\documentclass[iop,apj,useAMS,usenatbib]{emulateapj}
\usepackage{graphicx}
\usepackage{adjustbox,lipsum,natbib,epstopdf,color,amsmath,epsfig,url}
\usepackage[super]{nth}
\usepackage{hyperref}

\usepackage{siunitx}
\hypersetup{colorlinks=true, citecolor=blue, filecolor=magenta,
  urlcolor=cyan, linkcolor=blue}
\begin{document}
\author{Farhanul Hasan$^{1,2}$, Alison F. Crocker$\altaffiliation$^{1}$}
\affil{$^1$ Department of Physics, Reed College, Portland, OR 97202}
\affil{$^2$ Department of Astronomy, New Mexico State University, Las Cruces, NM 88003}

\title{Galaxies and Supermassive Black Holes at $\MakeLowercase{z} \leqslant 0.1$: The Velocity Dispersion Function}

\newcommand{\Msun}{M$_{\odot}$}
\newcommand{\edd}{$L_{Edd}$}
\newcommand{\mbh}{$M_{BH}$}
\newcommand{\kms}{km s$^{-1}$}
\newcommand{\mpcdex}{Mpc$^{-3}$ dex$^{-1}$}
\newcommand{\vmax}{$1/V_{\mathrm{max}}$}
\newcommand{\mmpc}{M$_{\odot}$ Mpc$^{-3}$}
\newcommand{\mpc}{Mpc$^{-3}$}
\newcommand{\scomp}{$\sigma$-complete}
\newcommand{\slim}{$\sigma_{\mathrm{lim}}$}
\newcommand{\slims}{$\sigma_{\mathrm{spec,lim}}$}
\newcommand{\slimm}{$\sigma_{\mathrm{mod,lim}}$}
\newcommand{\mlim}{$M_{r,\mathrm{lim}}$}
\newcommand{\slimz}{$\sigma_{\mathrm{lim}}(z)$}
\newcommand{\sspec}{$\sigma_{\mathrm{spec}}$}
\newcommand{\smod}{$\sigma_{\mathrm{mod}}$}

\begin{abstract}



We study the distribution of central velocity dispersion, $\sigma$, for $>$100000 galaxies in the SDSS at $0.01 \leqslant z \leqslant 0.1$. We construct the velocity dispersion function (VDF) from samples complete for all $\sigma$, where galaxies with $\sigma$ greater than the $\sigma$-completeness limit of the SDSS spectroscopic survey are included. We compare two different $\sigma$ estimates; one based on SDSS spectroscopy (\sspec{}) and another on photometric estimates (\smod{}). The \sspec{} for our sample is systematically higher than \smod{} for all ranges of $\sigma$, implying that rotational velocity may affect \sspec{} measurements. The VDFs measured from these quantities are remarkably similar for lower $\sigma$ values, but the \smod{} VDF falls faster than the \sspec{} VDF at $\log \sigma \gtrsim 2.35$.  Very few galaxies are observed to have $\sigma \gtrsim 350$ \kms{}. Despite differences in sample selection and methods, our VDFs are in close agreement with previous determinations for the local universe, and our results confirm that complete sampling is necessary to accurately discern the shape of the VDF at all ranges. We also find that both late and early type galaxies have \sspec{}$ > $\smod{}, suggesting that the rotation component of most galaxies figure significantly into \sspec{} measurements. Early-type galaxies dominate the population of high $\sigma$ galaxies, while late-type galaxies dominate the low $\sigma$ statistic. Our results warrant a more thorough and cautious approach in using long-slit spectroscopy to derive the statistics of local galaxies. Higher quality photometric measurements will enable more accurate and less uncertain measurements of the \smod{} VDF, as described here. A follow-up paper uses the final samples from this work in conjunction with the \mbh-$\sigma$ relation to derive the $z\leqslant 0.1$ black hole mass function (BHMF).


\end{abstract}


\section{Introduction}

Supermassive black holes (SMBH) exist at the centers of most, if not all, massive galaxies \citep{Richstone98, FF2005,  KH13}. The mass of a central black hole is known to correlate well with fundamental properties of a galaxy, including bulge mass \citep{KR95, MH03}, bulge luminosity \citep{Magorrian98, Graham07ML}, and stellar velocity dispersion \citep{Ferrarese2000, Gebhardt2000, MM13}. The tightness of these correlations have led many to postulate that the growth of galaxies is intrinsically connected to the growth of their SMBHs. Consequently, understanding the effect of SMBHs is of vital importance to any successful theoretical or semi-analytical model of galaxy formation and evolution \citep{Shankar09models, Shankar13models, HB14, Aversa15}.

Statistical studies of galaxies can be used as robust probes into galaxy evolution. Distribution functions have been used to describe the universe's galaxy populations in terms of fundamental galaxy properties such as luminosity \citep[e.g.,][]{Blanton01, Bernardi03, MR14}, stellar mass \citep[e.g.,][]{PG08, Kelvin14, Weigel16}, and velocity dispersion \citep[e.g.,][]{Sheth03, Bernardi10, SZG17}, helping us characterize the population of galaxies in both the local universe \citep{Choi07, Baldry12} and at high redshifts \citep{Marchesini07, Poz10}. Distribution functions have also been extremely helpful in describing the population of SMBHs via the black hole mass function \citep[BHMF; e.g.,][]{Salucci99, YT02, Graham2007, Li11}. In a subsequent paper, we will use the velocity dispersion distribution function we derive here to estimate a new local BHMF.



A crucial link between simulations and observations is the dark matter halo mass distribution function. While both the luminosity function (LF) and stellar mass function (SMF) of galaxies have been used to trace dark matter halos \citep{Yang08, Yang13}, connecting the LF to DM halo mass is non-trivial and the SMF has shown little dependence on galaxy characteristics such as morphology, color, and redshift \citep{Calvi12, Calvi13}. Thus, neither the LF nor the SMF may be very strong tracers of properties of the DM halo. The central stellar velocity dispersion (velocity dispersion, or $\sigma$, hereafter), on the other hand, is a dynamical measurement which reflects the stellar kinematics governed by the central gravitational potential well of a galaxy, and does not suffer from photometric biases or systematic uncertainties in modeling stellar evolution \citep{Conroy09, Bernardi13}. The central velocity dispersion has been observed to be strongly correlated with DM halo mass, leading some to conclude that it may be the best observable parameter connecting galaxies to their DM halo mass \citep{Wake12, Zahid16}.

Large-scale surveys of galaxies such as the Sloan Digital Sky Survey \citep[SDSS; ][]{York2000, St2002, DR12, DR14} have probed galaxy populations in great detail, across cosmic time. As a result, statistical analyses have been performed on these large datasets to establish global distributions of galaxies in terms of observables. In particular, SDSS galaxies have been used in deriving the velocity dispersion function (VDF) of galaxies in the field population \citep{Sheth03, Mitchell05, Choi07, SZG17} and the cluster population \citep{Munari16, Sohn17}. 

Deriving meaningful statistical information from the VDF requires a sample that is complete and unbiased. Volume-limited samples are incomplete below a certain magnitude limit, so distribution functions derived from such samples may be biased by selection effects \citep{PG08, Weigel16, Zahid16}. SMFs have been derived from samples complete in stellar mass, $M_{\star}$, where the $M_{\star}$ completeness limit was parametrized as a function of redshift \citep[e.g,][]{Fontana06, Marchesini09, Poz10, Weigel16}. Recently, \cite{SZG17} outlined an approach to empirically determine a redshift-dependent $\sigma$-completeness limit in order to derive a  $\sigma$-complete sample for a robust measurement of the VDF from local quiescent SDSS galaxies. Here, we follow their approach to  generate a $\sigma$-complete sample from $\sim360000$ SDSS galaxies -- both star-forming and quiescent -- at $z \leqslant 0.1$. 

Spectroscopic measurements through a fixed aperture are not reliable estimates for the true velocity dispersion of a galaxy due to contamination from galactic rotation \citep{Taylor10, Bezanson11, vu13}. Reported $\sigma$ values may include both the true $\sigma$ and an additional contribution from the rotational velocity, which is worse if the galaxy is more rotation dominated, the aperture extends further into the galaxy, or the galaxy is more inclined. This inherent bias in the velocity dispersion measurements may lead to inaccuracies in the derived VDF. Thus, in this paper, we also use velocity dispersions estimated by the approach of \cite{Bezanson11} to infer the true velocity dispersion based on the virial theorem and photometric estimates. Comparison of the VDF generated from this method is important to characterize the systematic uncertainties still present in two of our best ways of experimentally determining the velocity dispersion for nearby galaxies.



We use our $\sigma$-complete sample to construct a VDF for $\gtrsim100000$ local galaxies at $0.01 \leqslant z \leqslant 0.1$ from the SDSS, extending Sohn et. al. \citeyearpar{SZG17}'s analysis for local quiescent galaxies. In a follow-up paper, we use a similar \vmax{} methodology to determine the black hole mass function from our \scomp{} sample (Hasan \& Crocker, in prep).


This paper is organized as follows: In Section~\ref{data}, we present the data used in this work, including spectroscopic and photometric data from SDSS and morphological galaxy classifications from the Galaxy Zoo project \citep{Lintott08, Lintott11}. Here, we also derive two different estimates of velocity dispersion. We account for selection effects by constructing a \scomp{} sample from $z \leqslant 0.1$ galaxies in the SDSS in Section~\ref{comp}. We derive the VDF for the two different velocity dispersion estimates in Section~\ref{vdfsec} -- for our complete sample and for early and late type galaxies separately -- and compare these results with past works. Our conclusions are presented in Section~\ref{conc}. Throughout this paper, we assume a standard $\Lambda$-CDM cosmology with $H_0 = 70$ \kms{} ~Mpc$^{-1}$, $\Omega_m=0.3$ and $\Omega_\Lambda=0.7$.

\section{Data} \label{data}

In order to construct a VDF of local galaxies, we use velocity dispersion measurements from SDSS spectroscopy, as well as photometric data. SDSS, which began routine operations from 2000 \citep{York2000, St2002, Strauss2002}, has collected spectra of over a million galaxies and 100000 quasars, and imaged about a third of the sky ($\sim14000$ square degrees) \citep{DR7, DR8, DR12, DR14}. 


\subsection{Photometric data}

We use the SDSS main galaxy sample \citep{Strauss2002}, which is a magnitude-limited sample with $r$-band Petrosian \citeyearpar{Petrosian76} magnitude, $r_p < 17.77$ mags, limited to $z \lesssim 0.3$. The main galaxy sample is taken from SDSS Data Release 12 \citep{DR12}. We restrict our sample of local galaxies to $z \leqslant 0.1$, which brings our sample size from $\sim950000$ to $\sim360000$. Unlike Sohn et. al. \citeyearpar{SZG17}, we make no cuts based on $D_{n} 4000$ index to separate quiescent galaxies from star-forming ones. To correct for redshifting of light relative to our rest-frame, we apply K-corrections \citep{OS68} to the $r$-band magnitudes. We use the $z=0$ K-correction from the NYU Value Added Galaxy Catalog \citep{Blanton2005}.  Hereafter, the K-corrected absolute $r$-band magnitude is referred to as $M_r$.

\subsection{Velocity Dispersions} \label{vdisps}

SDSS selects $\sim99.9$\% of $r_p < 17.77$ objects from its images as spectroscopic targets. SDSS spectroscopy is $\sim95$\% complete, and produces spectra which cover the wavelength range of $3500-9000$ \AA{} with a resolution of $R\sim1500$ at 5000 \AA{} \citep{Strauss2002}. In particular, we adopt velocity dispersion measurements from the Portsmouth Group \citep{Thomas2013}, which are accessible as publicly available value added catalogs\footnote{\url{http://www.sdss.org/dr14/spectro/galaxy_portsmouth/}}. The Portsmouth templates are capped at $\sigma$ = 420 km s$^{-1}$, so that is the upper limit of reliable measurements for our sample. Furthermore, measurements where $\sigma \lesssim70$ \kms{} have low signal-to-noise (S/N) ratio, so they are unreliable too. The $\sigma$ measurements from the Portsmouth group all have S/N$>$10 and the median uncertainty is 7 km s$^{-1}$. 

The SDSS velocity dispersions are based on spectra obtained from a $1.5^{{\prime}{\prime}}$ fiber. These dispersions are afterwards corrected to an aperture of $R_e$/8 ($R_e$ = $r$-band effective radius) by adopting the calibration of \cite{Cappellari06}:

\begin{equation}
\text{$\left(\frac{\sigma_{\mathrm{SDSS}}}{\sigma_e}\right) = \left(\frac{R_{\mathrm{SDSS}}}{R_{e}/8}\right)^{-0.066}$,}
\end{equation}

where $R_{\mathrm{SDSS}} = 1.5^{{\prime}{\prime}}$ is the SDSS aperture radius and $\sigma_e$ is the velocity dispersion measured within a radius $R_e$/8 from the center. The aperture corrections are  small enough that they do not have a major impact on our results, and our qualitative conclusions don't change if we adopt a slightly different radial dependence. Hereafter, we refer to these $\sigma_e$ as \sspec{}. 

The velocity dispersion estimate for many emission line galaxies are likely compromised by rotation, in which case the reported $\sigma$ values are overestimated. Hence, the presence of rotation in most galaxies introduces biases in a VDF measured from \sspec{}. To account for this issue, we infer another estimate of $\sigma$ based on the virial theorem and photometric estimates of mass and size ($R_e$). In particular, we follow \cite{Taylor10} to determine a modeled $\sigma$, \smod{}, for each galaxy:

\begin{equation} \label{vireq}
\text{\smod{} $= \sqrt{\frac{G M_{\star}}{K_v(n) R_e}},$}
\end{equation}

where $M_{\star}$ is the stellar mass, $n$ the S\'ersic index, and $K_v(n)$ a correction term accounting for the effects of structure on stellar dynamics, approximated by \citep{Bertin02}:

\begin{equation}
\text{$K_v(n) \simeq \frac{73.32}{10.465 + (n-0.94)^2} + 0.954.$}
\end{equation}

For this analysis, we obtained $M_{\star}$ estimates from the MPA-JHU DR8 catalogues \citep{Kauffmann03, Brinchmann04} and $n$ and $R_e$ estimates from the NYU VAGC. The effective radii are based on fits to azimuthally averaged light profiles and are equivalent to circularized effective radii. 




\begin{figure}[!]
\centering
\includegraphics[scale=0.18]{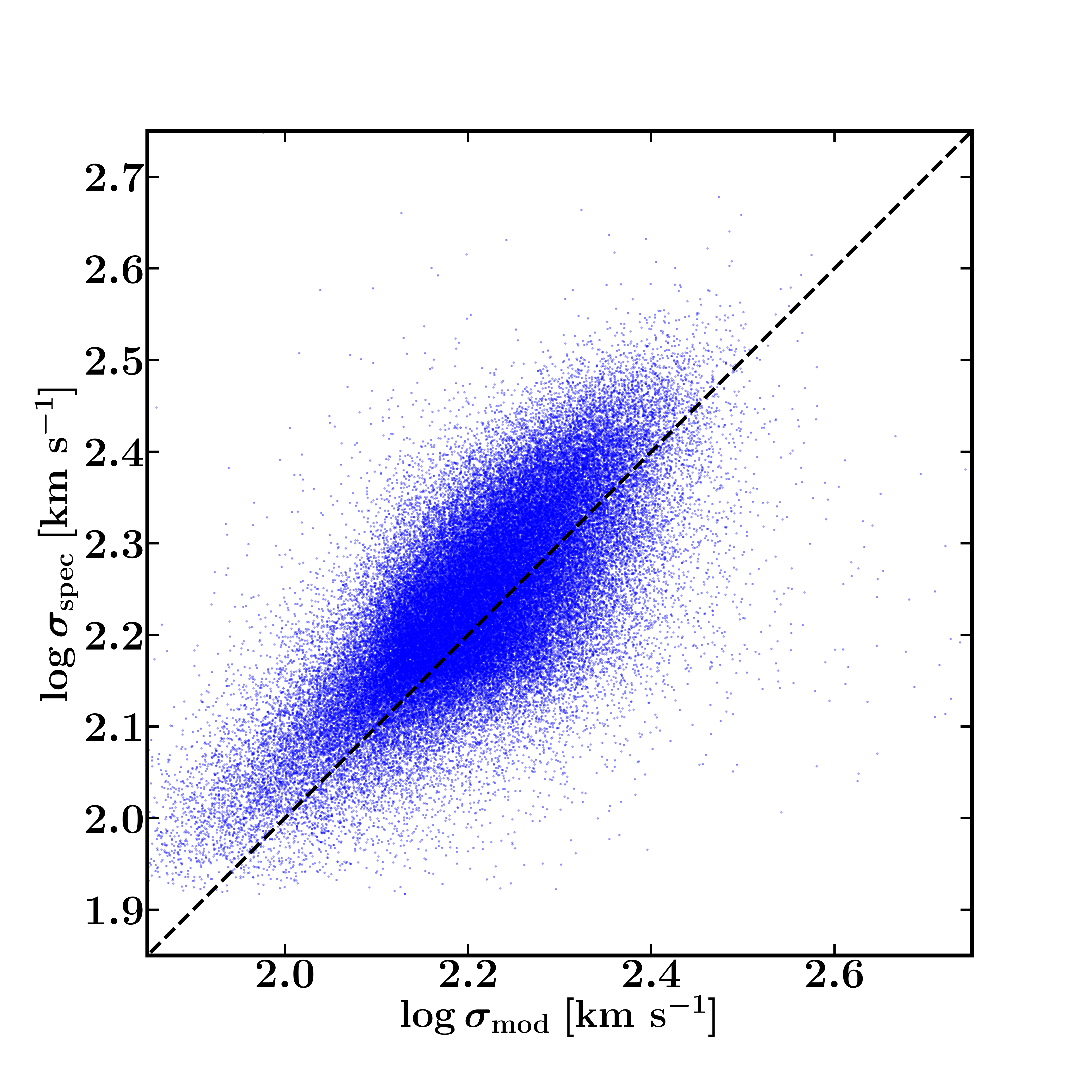}
\caption{Observed spectroscopic $\sigma$ (\sspec{}) vs. modeled $\sigma$ based on photometric estimates (\smod{}) for our final sampleThe dashed blue line indicates 1:1 match. Many $\log$ \smod{} values are below $\sim 1.9$, while few such values are observed for $\log$ \sspec{}.}
\label{specvmod}
\end{figure}

We compare the spectroscopically derived \sspec{} to the photometrically inferred \smod{} in Fig. \ref{specvmod}. We find that the galaxies in our sample have systematically higher \sspec{} than \smod{} for the entire range of velocity dispersions ($\log \sigma \sim 1.9-2.6$). In particular, we observe, like \cite{Bezanson11} and \cite{vu13}, that \smod{} is in general lower than \sspec{} for the low end (\smod{} $\lesssim 2$). The overestimate of \sspec{} relative to \smod{} is exactly what is expected for rotation-dominated galaxies which are more plentiful at lower velocity dispersions. Being noisy below $\sigma \lesssim70$ \kms{}, SDSS measurements don't report many galaxies with $\log \sigma \lesssim 1.9$. Seeing that a larger proportion of galaxies have $\log$ \smod{} $\lesssim 2$ than $\log$ \sspec{} $\lesssim 2$, \sspec{} may be giving us artificially high velocity dispersions for many galaxies. This leads to an incorrect representation of their true velocity dispersions. 

While the modeling adds a layer of complication to an otherwise directly measured quantity, we consider it a valuable complementary method of obtaining a velocity dispersion estimate. Going forward, we construct VDFs based on both these $\sigma$ estimates in section \ref{const}.

\subsection{Galaxy Zoo}

Galaxy Zoo is a web-based project in which the public is invited to visually classify galaxies from the SDSS by the two primary morphological types: spirals (late-types) and ellipticals and lenticulars (early-types) \citep{Lintott08}. The classifications of the general public agree with those of professional astronomers to an accuracy higher than 10 percent, and hundreds of thousands of systems are reliably classified by morphology at more than $>5\sigma$ confidence. A fundamental advantage of Galaxy Zoo is that it obtains morphological information by direct visual comparison instead of proxies such as color, concentration index or other structural parameters. Using these as proxies for morphology may introduce various systematic biases, leading to unreliable classifications and the need for more direct and robust means \citep{Lintott08, Bamford09}.

Each individual user assigns a vote for either spiral (late-type) or elliptical (early-type) for each galaxy. The fraction of the vote for each type for all objects is weighted as described in \cite{Lintott08} and then debiased in a consistent fashion (though \cite{Bamford09} outlines complications with the debiasing). Finally, a vote fraction of 80\% ensures an object is classified with certainty as either early or late type (the rest being classified ``uncertain"). Thus, we only selected objects with either spiral ($\sim15800$ galaxies) or elliptical ($\sim15400$ galaxies) flags.

\section{Selection Effects: Completeness of the sample} \label{comp}



A statistically complete sample is necessary to measure the statistical distribution of any galaxy property. A magnitude-limited sample such as the SDSS spectroscopic sample can easily be made volume-limited by choosing an appropriate absolute magnitude limit for each redshift $z$, such that within the entire volume all galaxies brighter than this magnitude limit should be detected. However, this volume-limited sample is only complete for a range in absolute magnitude, $M_r$, not for a range in velocity dispersions \citep{SZG17, ZG17}. Sohn et. al. (their Figure 1) show how there is substantial scatter in $\sigma$ for a fixed $M_r$. Similarly, we find that for example, at $M_{r} = -20.36$, \sspec{} varies from $\sim$50 km s$^{-1}$ to $\sim$315 km s$^{-1}$. Converting the absolute magnitude limit for a volume-limited
sample,\mlim{}, to a limit at which $\sigma$ is complete is therefore not a trivial exercise  \citep{Sheth03, SZG17}. 

Because of this broad distribution in \sspec{} and \smod{} for any $M_r$, there are many low $\sigma$ galaxies which make it to the sample by virtue of being bright enough, while some galaxies boasting high $\sigma$ are excluded for being less luminous. Completeness analysis helps us ensure that our sample isn’t preferentially selecting, say just the brightest galaxies. The goal in this section is to obtain the limit at which $\sigma$ is ($\sim95$\%) complete for any given volume (parametrized by redshift), for both the \sspec{} and \smod{} samples.

\subsection{Constructing $\sigma$-complete samples}
%
%

In order to generate a \scomp{} sample, we need to parametrize \slim{}, the limit at which the sample is complete for $\sigma > $ \slim{}, as a function of redshift $z$. We take an empirical approach in doing this, which closely follows the methodology of Sohn et. al. \citeyearpar{SZG17}. First, we divide the volume-limited sample of $0.01 \leqslant  z \leqslant 0.1$ into 40 volume-limited subsamples, each containing a volume with maximum redshift increased by $\bigtriangleup z = 0.0025$, starting from $z = 0.01$, and ending at $z = 0.1$. For each subsample, we derive  the \nth{95} percentile distribution of $\sigma$ for galaxies with $M_r \leqslant M_{r,\mathrm{lim}}$ to obtain \slim{}, the $\sigma$-completeness limit at the maximum redshift ($z_{\mathrm{max}}$) of the subsample. We repeat this process so that $\sigma_{\mathrm{lim}}$ is found for $z_{\mathrm{max}}$ ranging from 0.01 to 0.1.

%
%
%
%
%

\begin{figure}[htbp]
\centering
\includegraphics[scale=0.18]{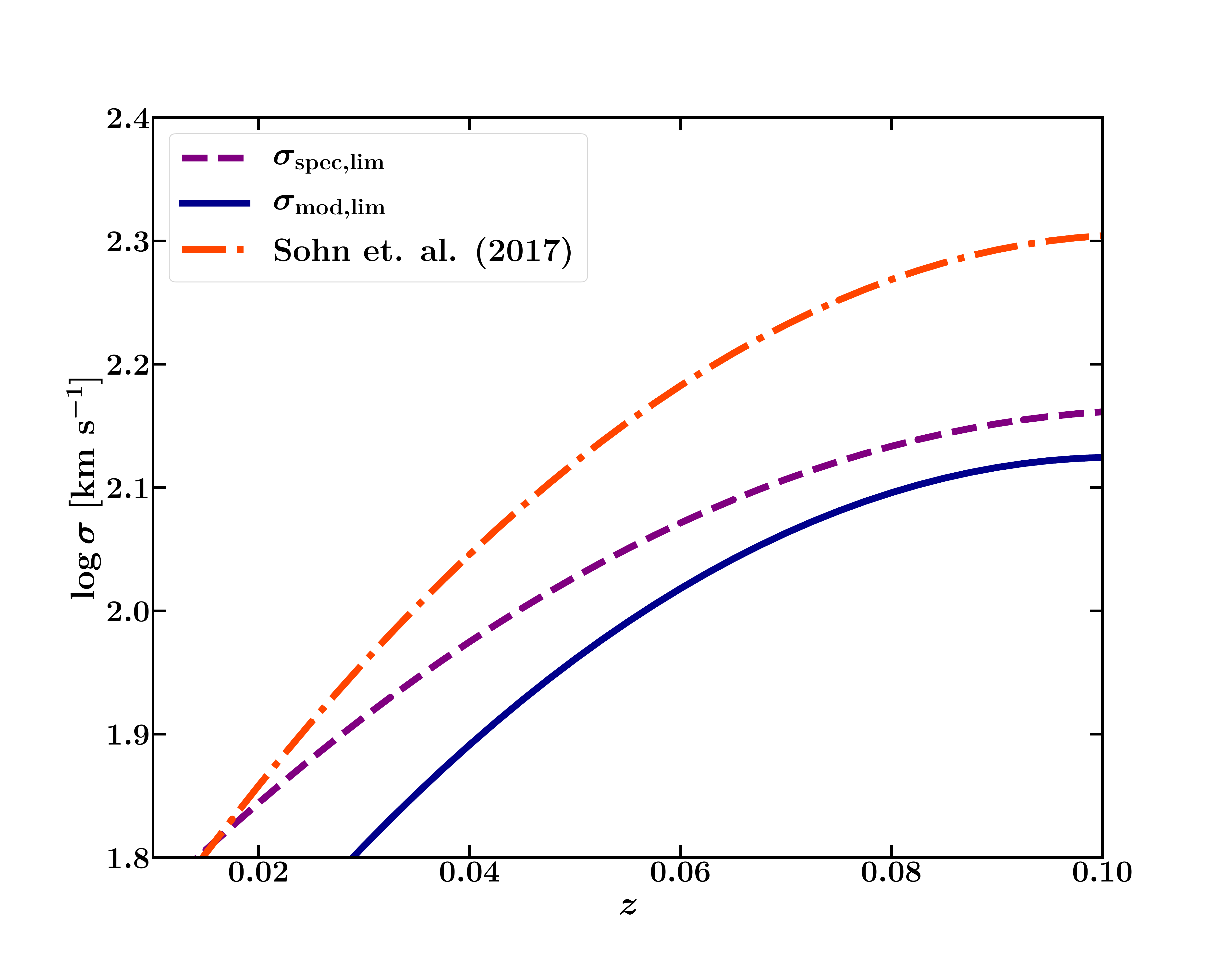}
\caption{$\sigma$-completeness limit as a function of redshift for the \sspec{} sample (dashed purple), \smod{} sample (solid dark blue), and Sohn et. al.'s \citeyearpar{SZG17} sample. All of these are \nth{2} order polynomial fits to the $\sigma$-completeness limit in bins of $\Delta z = 0.025$, with the Sohn et. al. fit for $0.03 \leqslant z \leqslant 0.1$, and both of our fits for $0.01 \leqslant z \leqslant 0.1$.}
\label{sigmalims}
\end{figure}

For both the \sspec{} and \smod{} samples, we fit the distribution of \slim{} against $z$ to a \nth{2} order polynomial. The polynomial fits we obtain are:

\begin{align} \label{slimzs}
\log \sigma_{\mathrm{spec, lim}}(z) = 1.68+ 9.10z - 42.83z^2 \\ \label{slimzm}
\log \sigma_{\mathrm{mod, lim}}(z) = 1.49 + 12.48z - 61.37z^2
\end{align}

The $\sigma$-completeness limit is plotted as function of redshift in Fig. \ref{sigmalims}, for our \sspec{} and \smod{} samples, as well as that of Sohn et. al. \citeyearpar{SZG17}, whose fit was also a \nth{2} order polynomial, but in the redshift range $0.03 \leqslant z \leqslant 0.1$. As expected, we find that \slim{} increases as $z_{\mathrm{max}}$ increases. We only include galaxies with \sspec{} $>$ \slims{} in the final \sspec{} sample and those with \smod{} $>$ \slimm{} in the final \smod{} sample. 

The final \sspec{} sample consists of {\boldmath $\sim118000$} galaxies and the final \smod{} sample contains {\boldmath $\sim105000$} galaxies in the redshift range $0.01 \leqslant z \leqslant 0.1$. These are both $\sim2.5$ times larger than Sohn et. al.'s sample, in which only quiescent galaxies were selected by the $D_{n} 4000$ index. About $83000$ galaxies were found in both samples. When comparing the properties of our \scomp{} samples with those of the magnitude-limited samples, we find that galaxies in the $\sigma$-complete samples are brighter and have less uncertain $\sigma$ measurements. The $\sigma$ values are also larger in general -- the mean \sspec{} and \smod{} increases from $\sim 100$ \kms{} to $\sim 165$ \kms{} and $\sim 75$ \kms{} to $\sim 160$ \kms{}, respectively, as we go from magnitude-limited samples to \scomp{} samples. The \scomp{} samples eliminate many of the low $\sigma$ galaxies which were present in the magnitude-limited sample, causing the mean $\sigma$ to go up.

\section{The Velocity Dispersion Function} \label{vdfsec}

The VDF is defined as the number density of galaxies with a given $\sigma$ per unit logarithmic $\sigma$. There are a few ways to compute the VDF, including the \vmax{} method \citep{Schmidt68}, the parametric maximum likelihood STY method \citep{Esf88}, and the non-parametric stepwise maximum likelihood (SWML) method \citep{Sandage79}. Several studies, including \cite{Weigel16} review these methods and the distribution functions resulting from them. We choose the \vmax{} method for our computation due to its simplicity and because we do not have to initially assume a functional form of the VDF. One drawback of this method is that it may produce biased results if there are inhomogeneities on large scales. The \vmax{} and SWML methods have been shown to produce equivalent results \citep{Weigel16, SZG17}.

\subsection{Constructing the VDF} \label{const}

The \vmax{} method takes into account the relative contribution to the VDF of each galaxy with dispersion $\sigma$, by volume-weighting the velocity dispersions. Each object is weighed by the maximum volume it could be detected in, given the redshift range and the $\sigma$-completeness of the sample. This corrects for the well-known Malmquist bias \citep{Malm25}. 


To generate our VDFs for both of our final samples, we first divide velocity dispersions in equal bins of width $\Delta \log \sigma = 0.02$, ranging from $\log \sigma = 1.8$ \kms{} to $\log \sigma = 2.7$ \kms{}. Then, the number density of galaxies in a specific $\sigma$ bin $j$ is given by the following sum:

\begin{equation} \label{vdfeq}
\text{$\Phi_{j}(\sigma) \Delta \log \sigma =  \sum_{i}^{N_{\mathrm{bin}}} \frac{1}{V_{\mathrm{max},i}}$,}
\end{equation}

where $N_{\mathrm{bin}}$ is the number of galaxies in the bin and \vmax{} is the maximum volume at which a galaxy $i$ with velocity dispersion $\sigma_i$ could be detected in. In a flat universe, the comoving volume, \vmax{}, is given by \citep{Hogg99}:

\begin{equation}
\text{$V_{\mathrm{max},i} = \frac{4\pi}{3} \frac{\Omega^{\mathrm{survey}}}{\Omega^{\mathrm{sky}}} \left[D_{C}(z_{\mathrm{max},i})^3 - D_{C}(z_{\mathrm{min},i})^3\right]$.}
\end{equation}

For all galaxies, $z_{\mathrm{min},i}$ = 0.03, since that is the lower redshift limit of our sample. $\Omega^{\mathrm{sky}} = 41253$ deg$^{\mathrm{2}}$ is the solid angle of the sky and $\Omega^{\mathrm{survey}} = 9200$ deg$^{\mathrm{2}}$ is the solid angle covered by SDSS. For $z_{\mathrm{max},i}$, we take the maximum redshift galaxy $i$ could have, based on $\sigma_i$ and the parametrized completeness limits from above (Fig. \ref{sigmalims} and eqs. \ref{slimzs} and \ref{slimzm}). 

The VDF, $\Phi(\sigma)$, was calculated from eq. \ref{vdfeq}, for both \sspec{} and \smod{}. The uncertainties on the VDF were estimated by a Monte Carlo method. We ran 10000 simulations of the VDF calculation, each time randomly modifying $\sigma$ values with the associated uncertainties, $\Delta \sigma$, assuming a gaussian error distribution. For \smod{}, we propagated the errors on uncertainties in $M_{\star}$, $R_e$, and $n$. The resulting VDF, and associated uncertainties, are tabulated in Table \ref{vdftab}. 

\begin{figure}[htbp]
\centering
\includegraphics[scale=0.15]{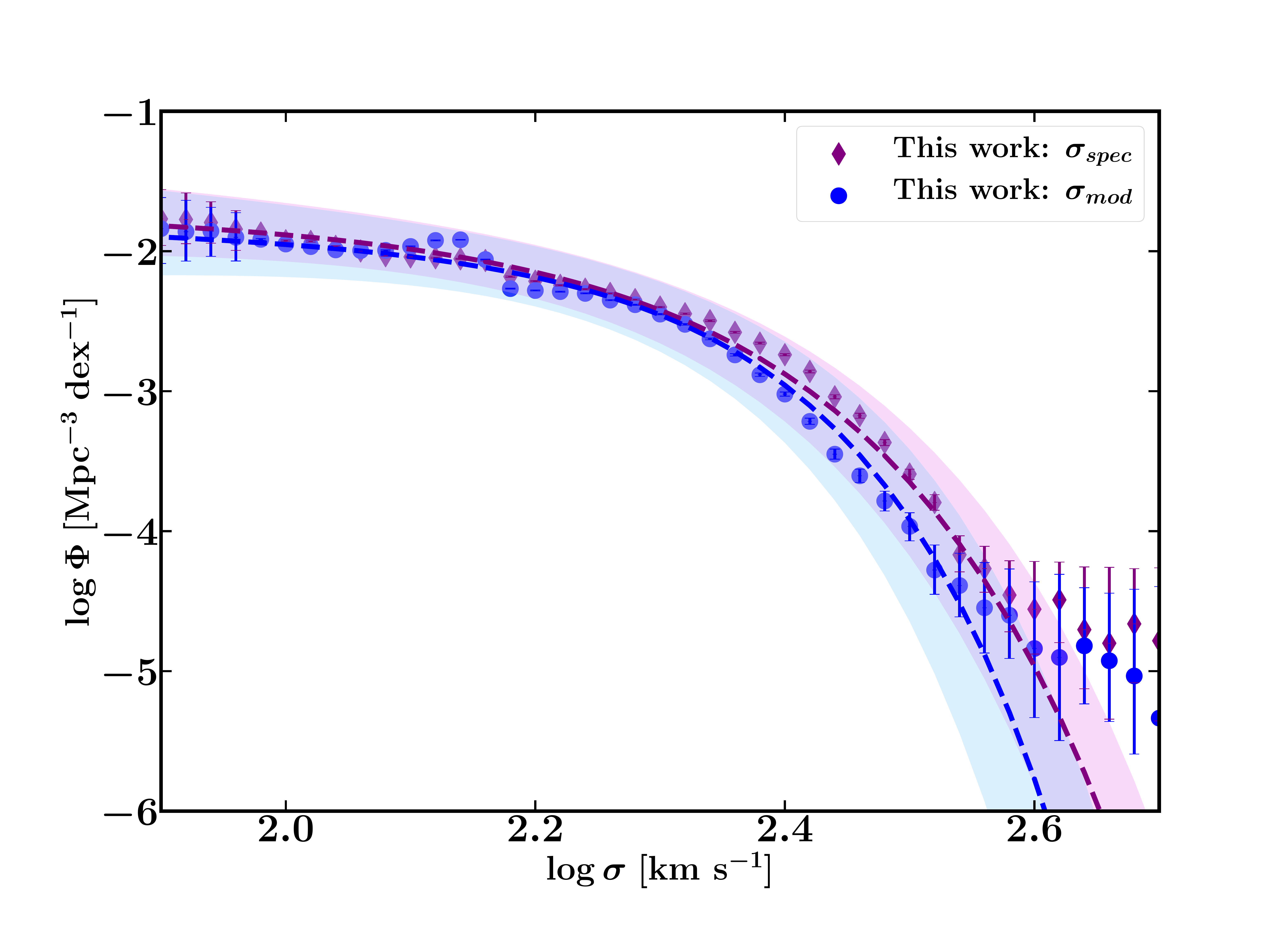}
\caption{VDFs for our $\sigma$-complete samples, based on \sspec{} -- purple diamonds -- and \smod{} -- blue circles. The purple dashed curve shows the 4-parameter best-fit Schechter function (eq. \ref{scheq}) for our \sspec{} VDF, while the blue dashed curve is the best-fit function for our \smod{} VDF. The pink and light blue shaded regions represent the 68\% confidence intervals for the \sspec{} and \smod{} fits, respectively.}
\label{vdf2}
\end{figure}

Fig. \ref{vdf2} shows the VDFs estimated from our \scomp{} \sspec{} and \smod{} samples. The VDFs are fairly reliable for $\sim1.9 \leqslant \log \sigma \leqslant 2.6$. There are very few objects with $\log \sigma \gtrsim 2.6$, resulting in uncertain VDF measurements for those bins. The shape of the VDFs are very similar throughout the range of velocity dispersions studied. Both VDFs decline slowly (toward higher $\sigma$) at low and medium $\sigma$ values ($\log \sigma \lesssim 2.3$), followed by an exponential fall-off at $\log \sigma \gtrsim 2.3$. The \smod{} VDF falls more rapidly than the \sspec{} VDF at $\log \sigma \gtrsim 2.35$, causing it to shift to the left relative to the \sspec{} VDF. At higher velocity dispersions (i.e. higher masses), we expect galaxies to be less rotation-dominated, which is in tension with our \smod{} VDF being lower than the \sspec{} VDF at $\log \sigma \gtrsim 2.5$. If anything, we would expect the \smod{} VDF to be lower than \sspec{} at the lowest $\sigma$s, or perhaps a systematic divergence between these throughout the $\sigma$ range studied. So we speculate that the calculated \smod{} VDF may be dubious, especially at higher $\sigma$.


We refrain from making claims about the ``true" VDF and which tracer -- \sspec{} or \smod{} -- does a better job of measuring it. Nevertheless, the \smod{} VDF is a valuable tool to interpret \sspec{} measurements, as well as the statistics based on SDSS spectroscopy. We note here that the apparent discontinuity just below $\log \sigma \sim 2.2$ observed in both the VDFs is likely not due to a real, physical phenomenon, but rather an artifact of our data analysis, and does not affect our main conclusions.

\begin{table*}
\centering \normalsize 
\caption{The \sspec{} VDF and \smod{} VDF with associated upper and lower limit uncertainties}

\begin{tabular}{ccccc} \toprule
\label{vdftab}

\boldmath{$\log \sigma$}  & \boldmath{$\Phi(\sigma_{\mathrm{spec}})$} & \boldmath{$\Delta \Phi(\sigma_{\mathrm{spec}})$} &  \boldmath{$\Phi(\sigma_{\mathrm{mod}})$} & \boldmath{$\Delta \Phi(\sigma_{\mathrm{mod}})$} \\ \footnotesize
 \boldmath{\bf [\kms{}]} & \footnotesize \boldmath{\bf [Mpc$^{-3}$ dex$^{-1}$]} & \footnotesize \boldmath{\bf [Mpc$^{-3}$ dex$^{-1}$]} & \footnotesize \boldmath{\bf [Mpc$^{-3}$ dex$^{-1}$]}  &  \footnotesize \boldmath{\bf [Mpc$^{-3}$ dex$^{-1}$]} \\
\hline \hline
\small 1.90 & \small \num[round-mode = figures,round-precision = 4]{0.017028354816005224} & $_{- \num[round-mode = figures,round-precision = 4]{0.20867565616851866}}^{+ \num[round-mode = figures,round-precision = 4]{0.18981613214474527}}$ & \small \num[round-mode = figures,round-precision = 4]{0.014484791408537041} & $_{- \num[round-mode = figures,round-precision = 4]{0.24871396458827671}}^{+ \num[round-mode = figures,round-precision = 4]{0.22277664695403046}}$ \\
\small 2.00 & \small \num[round-mode = figures,round-precision = 4]{0.011876225639561208} & $_{- \num[round-mode = figures,round-precision = 4]{0.005539931915383011}}^{+ \num[round-mode = figures,round-precision = 4]{0.005367064879843107}}$ & \small \num[round-mode = figures,round-precision = 4]{0.011251278004553208} & $_{- \num[round-mode = figures,round-precision = 4]{0.0051505384345179326}}^{+ \num[round-mode = figures,round-precision = 4]{0.006050829161923647}}$ \\
\small 2.06 & \small \num[round-mode = figures,round-precision = 4]{0.010043490431131737} & $_{- \num[round-mode = figures,round-precision = 4]{0.005012100727298547}}^{+ \num[round-mode = figures,round-precision = 4]{0.0045487347535804155}}$ & \small \num[round-mode = figures,round-precision = 4]{0.010101983526850825} & $_{- \num[round-mode = figures,round-precision = 4]{0.004388865727629055}}^{+ \num[round-mode = figures,round-precision = 4]{0.004562760730754095}}$ \\
\small 2.10 & \small \num[round-mode = figures,round-precision = 4]{0.009111182618836106} & $_{- \num[round-mode = figures,round-precision = 4]{0.004216494367598332}}^{+ \num[round-mode = figures,round-precision = 4]{0.004764258494464585}}$ & \small \num[round-mode = figures,round-precision = 4]{0.010815668668530353} & $_{- \num[round-mode = figures,round-precision = 4]{0.0035135026915097153}}^{+ \num[round-mode = figures,round-precision = 4]{0.0037254005439980014}}$ \\
\small 2.16 & \small \num[round-mode = figures,round-precision = 4]{0.008578261978107758} & $_{- \num[round-mode = figures,round-precision = 4]{0.003455766558100775}}^{+ \num[round-mode = figures,round-precision = 4]{0.0038864049494215}}$ & \small \num[round-mode = figures,round-precision = 4]{0.008721114049254413} & $_{- \num[round-mode = figures,round-precision = 4]{0.0028995006164695457}}^{+ \num[round-mode = figures,round-precision = 4]{0.002968607245300018}}$ \\
\small 2.20 & \small \num[round-mode = figures,round-precision = 4]{0.006105330195411446} & $_{- \num[round-mode = figures,round-precision = 4]{0.004375676373193428}}^{+ \num[round-mode = figures,round-precision = 4]{0.004349229492182617}}$ & \small \num[round-mode = figures,round-precision = 4]{0.0052401116583213805} & $_{- \num[round-mode = figures,round-precision = 4]{0.003449715580166128}}^{+ \num[round-mode = figures,round-precision = 4]{0.0030673848477221167}}$ \\
\small 2.26 & \small \num[round-mode = figures,round-precision = 4]{0.004992152687326059} & $_{- \num[round-mode = figures,round-precision = 4]{0.0046197758943672925}}^{+ \num[round-mode = figures,round-precision = 4]{0.004498991721751221}}$ & \small \num[round-mode = figures,round-precision = 4]{0.00447051559674813} & $_{- \num[round-mode = figures,round-precision = 4]{0.003713232040123877}}^{+ \num[round-mode = figures,round-precision = 4]{0.003969317008408295}}$ \\
\small 2.30 & \small \num[round-mode = figures,round-precision = 4]{0.003989765357770496} & $_{- \num[round-mode = figures,round-precision = 4]{0.004842043051580733}}^{+ \num[round-mode = figures,round-precision = 4]{0.0052031658445193295}}$ & \small \num[round-mode = figures,round-precision = 4]{0.0035518803664383288} & $_{- \num[round-mode = figures,round-precision = 4]{0.004431860378759695}}^{+ \num[round-mode = figures,round-precision = 4]{0.004847649461567026}}$ \\
\small 2.36 & \small \num[round-mode = figures,round-precision = 4]{0.002633245136128527} & $_{- \num[round-mode = figures,round-precision = 4]{0.006165424250809218}}^{+ \num[round-mode = figures,round-precision = 4]{0.005984175425752918}}$ & \small \num[round-mode = figures,round-precision = 4]{0.0018161675694710654} & $_{- \num[round-mode = figures,round-precision = 4]{0.008509803921568622}}^{+ \num[round-mode = figures,round-precision = 4]{0.008352214960057981}}$ \\
\small 2.40 & \small \num[round-mode = figures,round-precision = 4]{0.0018214432922582} & $_{- \num[round-mode = figures,round-precision = 4]{0.007443800432361017}}^{+ \num[round-mode = figures,round-precision = 4]{0.007580828573726}}$ & \small \num[round-mode = figures,round-precision = 4]{0.0009529274284261762} & $_{- \num[round-mode = figures,round-precision = 4]{0.014116262975778462}}^{+ \num[round-mode = figures,round-precision = 4]{0.015017301038062428}}$ \\
\small 2.46 & \small \num[round-mode = figures,round-precision = 4]{0.0006686978632693028} & $_{- \num[round-mode = figures,round-precision = 4]{0.01736252160899218}}^{+ \num[round-mode = figures,round-precision = 4]{0.01853979389119423}}$ & \small \num[round-mode = figures,round-precision = 4]{0.00024861843634371516} & $_{- \num[round-mode = figures,round-precision = 4]{0.04835013262599454}}^{+ \num[round-mode = figures,round-precision = 4]{0.04719893899204224}}$ \\
\small 2.50 & \small \num[round-mode = figures,round-precision = 4]{0.00025587255517602514} & $_{- \num[round-mode = figures,round-precision = 4]{0.034854226804123846}}^{+ \num[round-mode = figures,round-precision = 4]{0.038030927835051444}}$ & \small \num[round-mode = figures,round-precision = 4]{0.00010815231713625807} & $_{- \num[round-mode = figures,round-precision = 4]{0.10320731707317096}}^{+ \num[round-mode = figures,round-precision = 4]{0.09791463414634062}}$ \\
\small 2.60 & \small \num[round-mode = figures,round-precision = 4]{2.7697544632456338e-05} & $_{- \num[round-mode = figures,round-precision = 4]{0.3409999999999986}}^{+ \num[round-mode = figures,round-precision = 4]{0.3203333333333346}}$ & \small \num[round-mode = figures,round-precision = 4]{1.4508237664619986e-05} & $_{- \num[round-mode = figures,round-precision = 4]{0.49318181818181794}}^{+ \num[round-mode = figures,round-precision = 4]{0.47661090909090953}}$ \\
\hline \\

\end{tabular}
\tablecomments{Table 1 will be published in its entirety in machine-readable format when we submit to the Astronomical or Astrophysical Journal}
\end{table*}

\subsection{Schechter function}

The LF, SMF, VDF and other distribution functions of galaxies are often described by a Schechter \citeyearpar{Sch76} function. Schechter's original function to approximate the observed LF, $\Phi(L)$, exhibits a power-law behavior at lower values of $L$, followed by an exponential cutoff at a characteristic $L_{\star}$. We adopt a similar expression for our VDF, $\Phi(\sigma)$, but find that adding an additional parameter, $\beta$, gives a better approximation to the data. Thus, following the functional form of the BHMF from \cite{AR2002}, we adopt the following modified Schechter function:

\begin{equation} \label{scheq}
\text{$\Phi(\sigma) ~d\sigma=  \Phi_{\star} \left(\frac{\sigma}{\sigma_{\star}}\right)^{\alpha+1} \exp \left[1- \left(\frac{-\sigma}{\sigma_{\star}}\right)^\beta\right] ~d\sigma$.}
\end{equation}

where the characteristic truncation value of $\sigma$ is $\sigma_{\star}$ and $\Phi(\sigma_{\star}) =  \Phi_{\star}$. The slope of the power law is $\alpha+1$ ($\alpha = -1$ corresponds to a flat distribution at $\sigma < \sigma_{\star}$). The \sspec{} and \smod{} VDFs are both parametrized with the form of eq. \ref{scheq}. The parameters of these fits  are given in Table \ref{schtab}. The uncertainties on the parameters were derived by using the same Monte-Carlo simulations by which uncertainties on the VDF were found (assuming gaussian error distributions again).

\begin{table*}[htbp] 
\begin{center}
\normalsize 
\label{schtab}
 \caption{Best-fit parameters for modified Schechter function (eq. \ref{scheq}) fits to VDFs for our \sspec{} and \smod{} sample}
\begin{tabular}{ c | c | c | c | c } \toprule

{\bf Sample}   & \boldmath${\alpha}$ & \boldmath${\beta}$ & \boldmath${\sigma_{\star}}$ &  \boldmath${\Phi_{\star}}$ \\
 & &  & \boldmath{\bf [\kms{}]} & \boldmath{\bf [\mpcdex]} \\
\hline \hline
\sspec{}  & $-1.15 \pm 0.31$    & $2.35 \pm 0.31$           & $172.2 \pm 23$     & $(5.86 \pm 0.35) \times 10^{-3}$  \\ 
\hline 
\smod{}  & $-1.21 \pm 0.28$         & $2.95 \pm 0.37$           & $189.1 \pm 27$     & $(4.20 \pm 0.29) \times 10^{-3}$  \\ 
\hline \hline
 
\end{tabular}

\end{center}
\end{table*}

\normalsize

\subsection{Comparison with literature}

\begin{figure*}[htbp]
\centering
\includegraphics[scale=0.23]{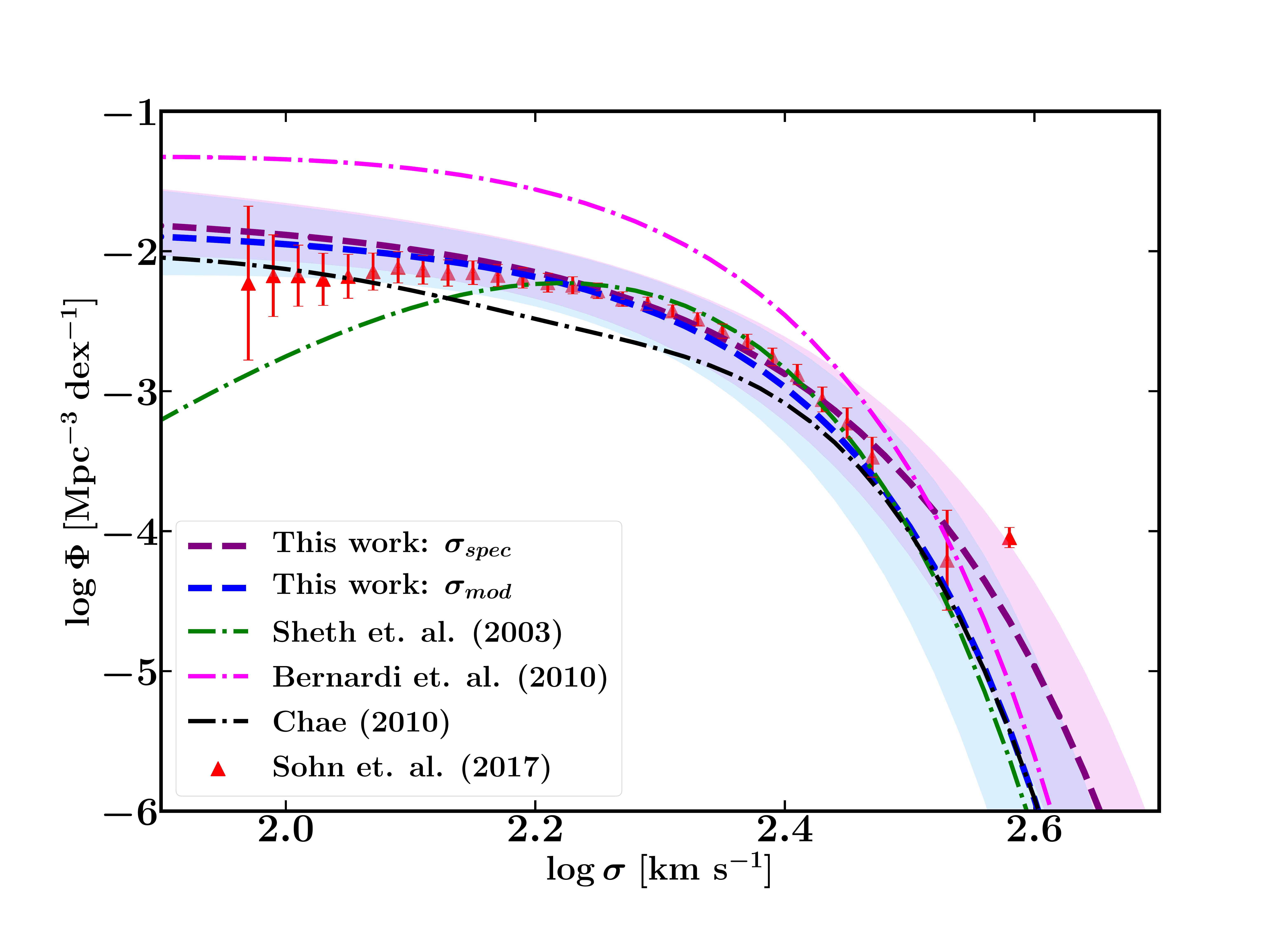}
\caption{Comparison of the VDFs from this work with with past works. Our \sspec{} VDF is the dashed purple curve ($\pm1\sigma$ region in pink), and our \smod{} VDF is shown in dashed blue ($\pm1\sigma$ region in light blue). The past VDFs are shown as dot-dashed curves: \cite{Sheth03} -- green; \cite{Bernardi10} -- magenta; \cite{Chae10} -- black, and individual data points: Sohn et. al. \citeyearpar{SZG17} -- red triangles.}
\label{vdfcomp}
\end{figure*}


The differences in the shape of the derived VDF for low $\sigma$, and in the characteristic $\sigma$ at which the VDF truncates, were attributed by Choi et. al. \citeyearpar{Choi07} to differences in sample selection. For example, Sheth et. al.'s \citeyearpar{Sheth03} sample of only early-type galaxies, was not complete in $\sigma$, and their resulting VDF declined noticeably for $\log \sigma \lesssim 2.2$ \kms{}. Comparison between previous determinations of the VDF illuminate the fact that astute sample selection is critical in determining the VDF \citep{Choi07}. 


In Fig. \ref{vdfcomp}, we compare our VDFs derived from the \scomp{} sample with previous determinations. Of these, the Bernardi et. al. \citeyearpar{Bernardi10} and Sohn et. al.  VDFs were determined directly from spectroscopic measurements of $\sigma$ from SDSS galaxies. Chae \citeyearpar{Chae10} used a combination of Monte Carlo-realized SDSS early-type and late-type VDF to estimate the total VDF. Of the VDFs shown, the Sheth et. al. \citeyearpar{Sheth03} VDF is an early-type VDF only. Furthermore, Sohn et. al.'s sample selection criteria essentially limited their sample to early-type galaxies since the 4000 \AA{} break strength is strongly correlated with stellar population age, and by extension, galaxy type \citep{Kauffmann03, Zahid16}.


The shapes of both of our VDFs are very similar to those of Sohn et. al., Bernardi et. al. and Chae et. al., despite differences in methods and sample selection. The Sohn et. al. sample is complete in $\sigma$ like ours and is a reliable estimate of the quiescent population VDF. Their VDF flattens somewhat at $\log \sigma < 2.1$, whereas both of ours exhibit a slow upturn towards lower $\sigma$s. The Bernardi et. al. VDF, which has a similar shape to ours (albeit predicts much higher number densities), drew from a sample which was magnitude-limited after correction for incompleteness. They used \vmax{} like us, in contrast to Sohn et. al. (who used the SWML method, but found nearly equivalent results with \vmax{}). Chae, on the other hand, converted the LFs of early-type and late-type galaxies (from a magnitude-limited sample) to VDFs separately and added a correction term for high $\sigma$ galaxies.


It is certainly interesting that the VDF from an inferred quantity -- \smod{} -- reproduces results obtained from direct measurements of the velocity dispersion across the literature. The approximate match between our \smod{} VDF and \sspec{} VDFs from a variety of sample selection methods is a promising sign for better uncovering the true velocity dispersion statistics for large samples of galaxies. \cite{Bezanson11} derived the total VDF for $0.3 < z \leqslant 1.5$ based on \smod{} (eq. \ref{vireq}) and found that the VDF flattens toward lower $\sigma$s with decreasing redshift, and that the number of galaxies at the highest $\sigma$ bins ($\log \sigma \gtrsim 2.5$) change little. We find that their $0.3 < z \leqslant 0.6$ \smod{} VDF is similar to ours, possibly implying little evolution from $z\sim0$ to $z\sim0.3$. In any case, the agreement of our \smod{} VDF with our \sspec{} VDF as well as those in the literature, compels a more thorough assessment of the differences in $\sigma$ estimates from spectroscopic and photometric measurements, especially so that the effect of galactic rotation is taken into account in correcting the observed \sspec{} and obtaining the ``true" estimate of central velocity dispersion.


\subsection{Early and late types}

\begin{figure}[htbp]
\centering
\includegraphics[width=8.75cm,height=5cm]{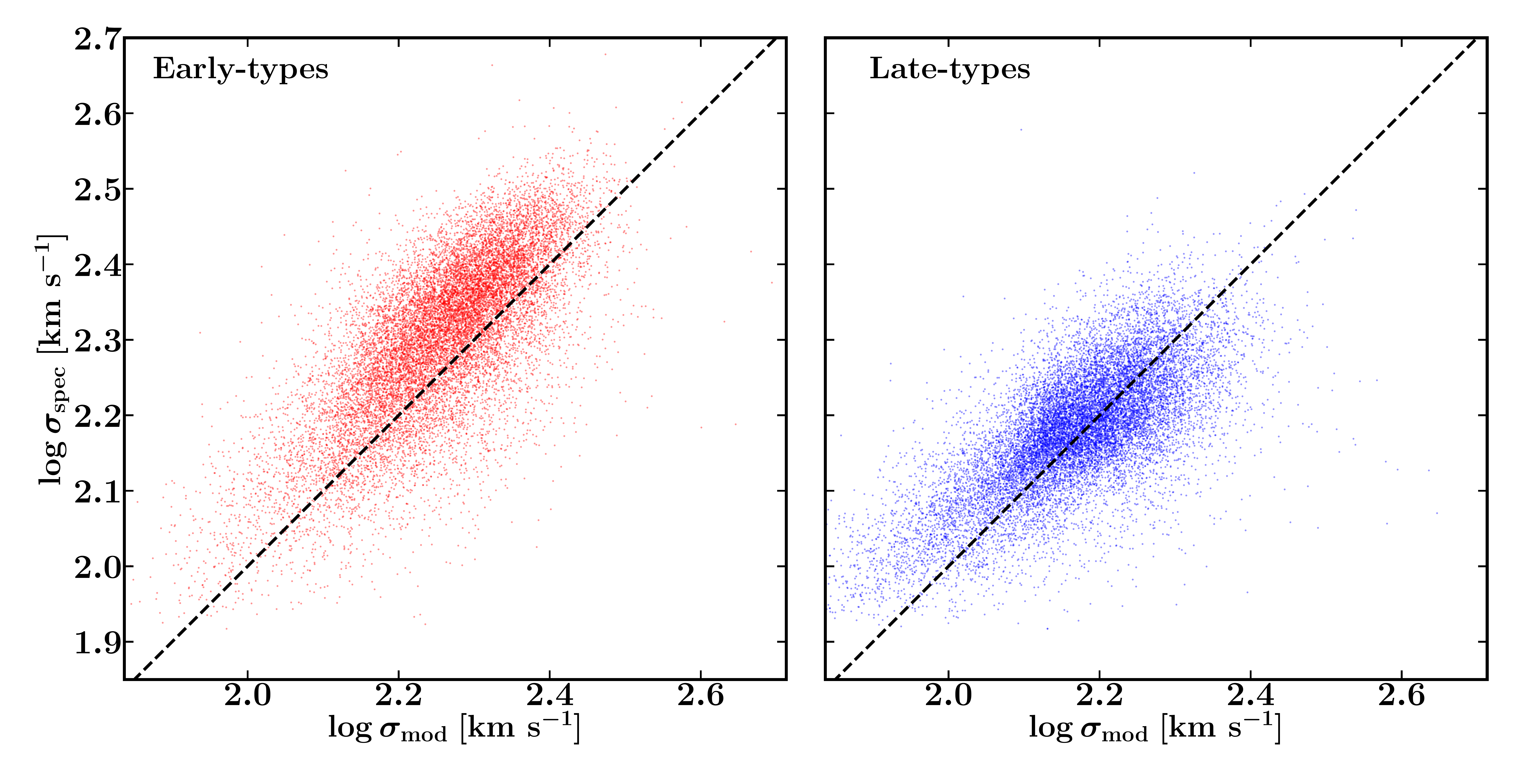}
\caption{Comparison of velocity dispersion estimates for early-type galaxies (left) and late-types galaxies (right), similar to Fig. \ref{specvmod}.}
\label{stypes}
\end{figure}

\begin{figure}[htbp]
\centering
\includegraphics[width=8.75cm,height=5cm]{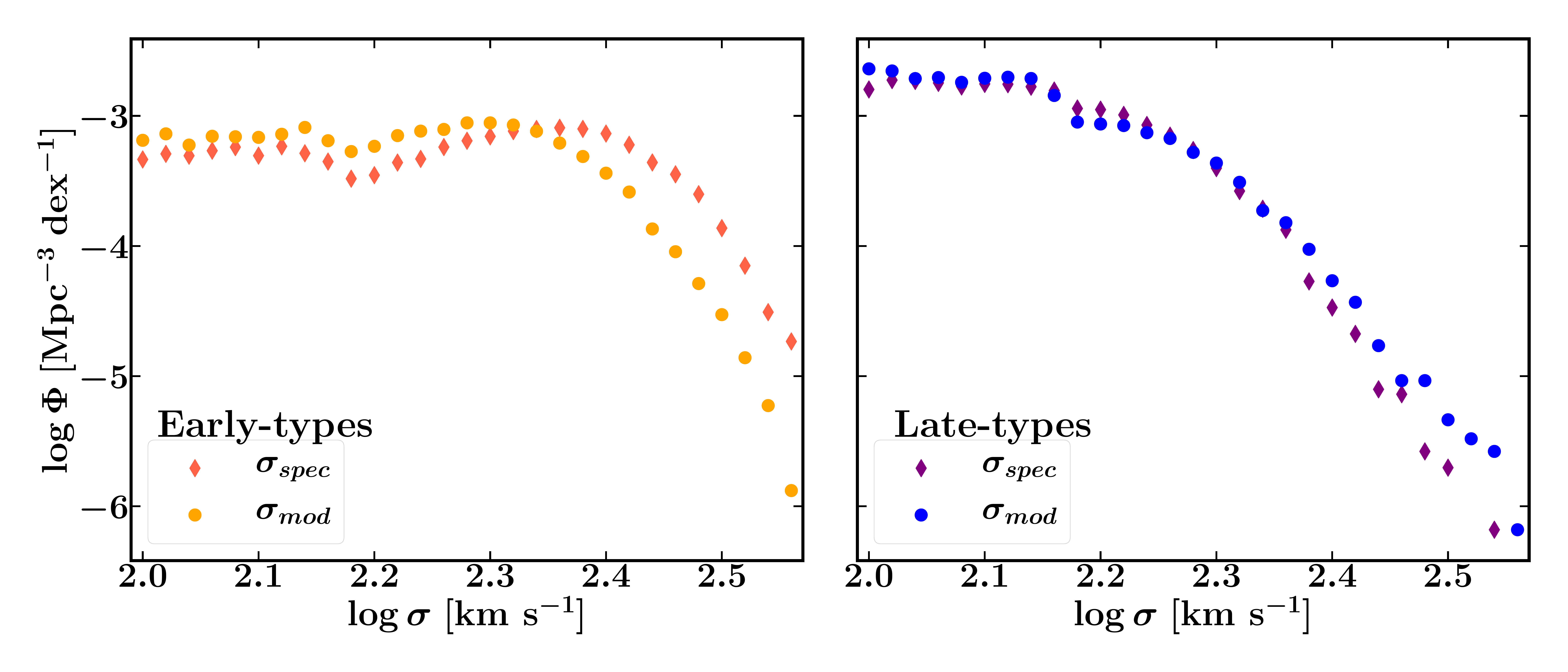}
\caption{VDF for early-type galaxies (left) and late-types galaxies (right). {\it Left}: red = \sspec{} VDF, orange = \smod{} VDF.  {\it Right}: purple = \sspec{} VDF, blue = \smod{} VDF. }
\label{vdftypes}
\end{figure}


Using Galaxy Zoo's classification of $\sim32000$ objects as either early or late type galaxies, we repeated the above analysis for both subsets separately, based on both \sspec{} and \smod{}. Early type galaxies, comprising of ellipticals and lenticulars, tend to be the most massive galaxies in the universe. These are populated by older, redder stars and due to a lack of cold gas reservoirs, have low star formation activity \citep{Jura77, Crocker11}. In contrast, spirals, or late type galaxies, tend to be less massive, bluer and populated mostly by younger stars. They have been shown to exhibit much higher rates of star formation than early type systems \citep{Young96, Seigar17}.


While no type-dependent separation of the VDF based on modeled velocity dispersions can be found in the literature yet, many studies have examined the distribution of early and late types based on the spectroscopic $\sigma$. Bernardi et. al. \citeyearpar{Bernardi10}, for example, derived the VDF of low redshift SDSS galaxies (based on \sspec{}) and found that the VDF obtained from E+S0s (early-types) declined toward low $\sigma$, but the addition of spirals like Sas increased the number density at those ranges. In a similar way, Weigel et. al. \citeyearpar{Weigel16} observed a declining SMF for lower masses for local early-types in the SDSS, while the late-type SMF showed an increasing SMF for lower masses.

Fig. \ref{stypes} compares the \sspec{} and \smod{}, as in Fig. \ref{specvmod}, for both galaxy types. While it would be expected that spirals in general have \sspec{}$ > $\smod{}, we see the same is true for early-types in our sample. In fact, the early-type \sspec{} values seems to be higher than \smod{}, compared to late-types. For the late-types, we find that \smod{} predicts much lower $\sigma$ values than \sspec{} for the lowest velocity dispersions ($\log \sigma \lesssim 2$). The offset in both plots suggests that \sspec{} may need to be corrected for rotation in not just late-type galaxies, but early-types too. \cite{BM09} note that lenticulars (S0) are mostly fast rotators, and that even most giant ellipticals may rotate appreciably.



As Fig. \ref{vdftypes} shows, the VDFs from our early-type and late-types galaxies do differ. As expected, late-type galaxies have a high fraction of low $\sigma$ galaxies and contribute the most to the low-$\sigma$ distribution, while early-type galaxies dominate the high-$\sigma$ distribution. Here, we observe a flattening of the early-type VDF toward low \smod{}, and a slight rise in the late-type VDF for $\log \sigma_{\mathrm{mod}} \lesssim 2.15$. Both the \sspec{} VDFs seem to flatten at the lowest $\sigma$s. For the early-types, we see that the \sspec{} VDFs overestimate the \smod{} VDFs at $\log \sigma \gtrsim 2.4$, akin to the VDFs for our full samples in Fig. \ref{vdf2}, but the converse is true for the late-types. Interestingly enough, it appears that the early-type VDFs could be approximated by a double Schechter function. However, we restrain ourselves from arriving at major conclusions from the early and late type sample alone, as they represent only about a third of our final samples and could therefore be biased in some form.

\section{Conclusion} \label{conc}

We measure the VDF of galaxies from the SDSS at $0.01 \leqslant z \leqslant 0.1$, using two different estimates of velocity dispersion: the directly measured \sspec{}, and the photometrically inferred \smod{}. We observe systematically higher \sspec{} relative to \smod{} in these galaxies, possibly due to SDSS spectroscopy being unable to separate the rotation component from the ``true" velocity dispersion. We construct \scomp{} samples following the approach of Sohn et. al. \citeyearpar{SZG17}, by selecting every galaxy with $\sigma > $ \slimz{}, the $\sigma$-completeness limit as a function of redshift for an originally magnitude-limited sample at $r_p < 17.77$. Our final \sspec{} and \smod{} samples consist of over 100000 galaxies each and is complete for all $\log$ \sspec{} $\gtrsim 1.8$ and $\log$ \smod{} $\gtrsim 1.6$, respectively. These represent the largest samples ever used in measureing the $z \leqslant 0.1$ VDF.


Our measured VDFs decline slowly toward higher $\sigma$ at low $\sigma$, followed by an exponential decline at high $\sigma$. The most uncertain number densities are observed for the highest $\sigma$ bins, since there are very few galaxies with $\log \sigma \gtrsim 2.6$ \kms{}. The VDFs derived from \sspec{} and \smod{} agree very well, especially at $\log \sigma \lesssim 2.3$, but the \smod{} VDF falls faster than the \sspec{} VDF at $\log \sigma \gtrsim 2.35$. This difference may be caused by our \smod{} VDF being erroneous at the high $\sigma$ range, so we caution against immediately drawing conclusions about which tracer is a better descriptor of the true VDF, or about the extent to which the \sspec{} VDF overestimates the true VDF. However, the fact that our results do show divergences between \sspec{} and \smod{} -- hypothetically the same physical property -- and their respective VDFs supports the argument that values of $\sigma$ measured from SDSS long-slit spectroscopy may be biased in some form.




We also use Galaxy Zoo's classification of a fraction of galaxies in our sample as either early-type or late-type, and obtain VDFs for both subsamples separately. We see that \sspec{} is in general higher than \smod{}, especially at $\log \sigma \lesssim 2$, for both types, again implying rotational velocity could be contaminating \sspec{}. Overall, early-type galaxies dominate the VDF at high $\sigma$ ($\log \sigma \gtrsim 2.4$), while late-types dominate the distribution at lower values ($\log \sigma \lesssim 2.1$). The \sspec{} VDF predicts higher number densities at $\log \sigma \gtrsim 2.4$ for early-types, while the \smod{} VDF is higher at that range for the late-types. 


A comparison with the local VDFs derived in the literature from direct spectroscopic measurements agree remarkably well with both our \smod{} and \sspec{} VDFs, despite differences in sample selection and methods. Thus, we believe that the literature and our work has narrowed down on the ``true" VDF for $z\leqslant0.1$ galaxies. In this work, we've offered new perspectives to correct for the uncertainties that arise in such a measurement.

 

We would like to stress that there are both advantages and disadvantages of using photometric estimates to measure velocity dispersion and the VDF. Direct studies of the VDF at higher redshifts are challenging because it is difficult to obtain kinematic measurements of thousands of galaxies, while photometric properties like mass and radius are more readily available for larger samples. This enables VDF measurements at higher redshifts using inferred velocity dispersions. However, most measurements of galaxy sizes are from ground-based observations, which suffer from significant uncertainties. Future space-based imaging would surely improve the accuracy of photometric measurements such as galaxy radii and S\'ersic indices, leading to more accurate estimates of \smod{}.

The VDF may be an important tool in probing the mass distribution of DM halos, and its accurate reproduction by numerical simulations is paramount in tightening our constraints on the evolution of galaxies. Combining VDF estimates from large samples such as the one used here with numerical simulations would therefore be a powerful window to understanding the large-scale structure formation and evolution of the universe. A follow-up paper will present the black hole mass function derived from our \scomp{} sample, using \cite{Remco16}'s scaling of $\sigma$ to SMBH mass, \mbh{}.

\section*{Acknowledgements}

F.H would like to thank Jubee Sohn for his helpful comments regarding the sample selection at the beginning of this research. This research has made use of NASA’s Astrophysics Data System (ADS) Bibliographic Services. This publication has also made use of Astropy, a community-developed core PYTHON package for Astronomy (Astropy Collaboration, 2013). Furthermore, we used the Tool for OPerations on Catalogues And Tables (TOPCAT\footnote{\url{http://www.starlink.ac.uk/topcat/}}) and AstroConda, a free Conda channel maintained by the Space Telescope Science Institute (STScI) in Baltimore, Maryland.

Galaxy Zoo is supported in part by a Jim Gray research grant from Microsoft, and by a grant from The Leverhulme Trust. Galaxy Zoo was made possible by the involvement of hundreds of thousands of volunteer ``citizen scientists."

Funding for SDSS-III has been provided by the Alfred P. Sloan Foundation, the Participating Institutions, the National Science Foundation, and the U.S. Department of Energy Office of Science. SDSS-III is managed by the Astrophysical Research Consortium for the Participating Institutions of the SDSS-III Collaboration. The Participating Institutions are the American Museum of Natural History, Astrophysical Institute Potsdam, University of Basel, University of Cambridge, Case Western Reserve University, University of Chicago, Drexel University, Fermilab, the Institute for Advanced Study, the Japan Participation Group, Johns Hopkins University, the Joint Institute for Nuclear Astrophysics, the Kavli Institute for Particle Astrophysics and Cosmology, the Korean Scientist Group, the Chinese Academy of Sciences (LAMOST), Los Alamos National Laboratory, the Max Planck Institute for Astronomy (MPIA), the Max Planck Institute for Astrophysics (MPA), New Mexico State University, Ohio State University, University of Pittsburgh, University of Portsmouth, Princeton University, the United States Naval Observatory and the University of Washington. The SDSS website is \url{www.sdss.org/}.

\bibliographystyle{astron}
\bibliography{Refs.bib}

\end{document}